**Key Points:**

- Latitudinal distributions of solar protons reveal strong day-night asymmetries
- A simple applicable cutoff latitude parameterization is provided

**Correspondence to:**
H. Nesse Tyssøy,
hilde.nesse@ift.uib.no

**Citation:**
Nesse Tyssøy, H., and J. Stadsnes (2015), Cutoff latitude variation during solar proton events: Causes and consequences, *J. Geophys. Res. Space Physics*, *120*, 553–563, doi:10.1002/2014JA020508.

Received 13 AUG 2014
Accepted 18 DEC 2014
Accepted article online 22 DEC 2014
Published online 30 JAN 2015

# Cutoff latitude variation during solar proton events: Causes and consequences

**H. Nesse Tyssøy[1] and J. Stadsnes[1]**

[1]Birkeland Centre for Space Science, Department of Physics and Technology, University of Bergen, Bergen, Norway

**Abstract** To accurately quantify the effect of solar proton events (SPEs) on the atmosphere requires a good estimate of the particle energy deposition in the middle atmosphere (60–100 km) and how the energy is distributed globally. Protons in the energy range 1–20 MeV, depositing their energy in the middle atmosphere, are subject to more complex dynamics with strong day-night asymmetries compared to higher-energy particles. Our study targets six SPEs from 2003 to 2012. By using measurements from the Medium Energy Proton and Electron Detector on all available Polar Orbit Environment Satellites (POES), we show that in the main phase of geomagnetic storms the dayside cutoff latitudes are pushed poleward, while the nightside cutoff latitudes have the opposite response, resulting in strong day-night asymmetries in the energy deposition. These features cannot be measured by the frequently used Geostationary Operational Environmental Satellites (GOES). Assuming that the protons impact the polar atmosphere homogeneously above a fixed nominal latitude boundary will therefore give a significant overestimate of the energy deposited in the middle atmosphere during SPEs. We discuss the magnetospheric mechanisms responsible for the local time response in the cutoff latitudes and provide a simple applicable parameterization which includes both dayside and nightside cutoff latitude variability using only the *Dst*, the northward component of the interplanetary magnetic field, and solar wind pressure. The parameterization is utilized on the GOES particle fluxes, and the resulting energy deposition successfully captures the day-night asymmetry in good agreement with the energy deposition predicted from the POES measurement.

## 1. Introduction

On rare occasions, solar protons are accelerated into energies that can penetrate the Earth's closed magnetic field lines and impact the atmosphere at latitudes normally shielded from the direct solar wind impact. The term solar proton event (SPE) is usually applied to proton fluxes of more than 10 particles $cm^{-2} s^{-1} sr^{-1}$ with energies in excess of 10 MeV (http://umbra.nascom.nasa.gov/SEP/). The influence of SPEs upon the middle atmosphere has been widely studied through both measurements and models [e.g., *Jackman et al.*, 2001; *Funke et al.*, 2011; *Sinnhuber et al.*, 2012]. It has been thoroughly established that particle energy deposition in the atmosphere produce copious amount of chemically reactive nitrogen and hydrogen species, which reduce the ozone concentration and alter the radiative balance. Consequently, this could alter both the latitudinal temperature gradients and perturb the vertical energy transfer throughout the lower atmosphere [e.g., *Gray et al.*, 2010]. To accurately quantify the effect of SPEs on the atmosphere requires a good estimate of the energy deposited in the atmosphere and how the energy is distributed globally.

The access of solar protons into the Earth's magnetosphere is mainly controlled by the magnetospheric magnetic field [e.g., *Størmer*, 1955; *Smart and Shea*, 2001] and is limited in latitude by the particles' energy. The cutoff latitude is defined as the lowest latitude to which a solar proton of a given energy can penetrate [*Kress et al.*, 2010]. Both the strength and shape of the magnetospheric magnetic field depend on solar wind parameters through the generation of electrical current systems, such as the magnetopause current, ring current, and the cross-tail current. Simple schematics of the expected pattern of the magnetic field resulting from the different currents are shown in Figure 1. An intensification of the magnetopause currents will strengthen the dayside magnetic field, pushing the cutoff latitudes poleward. An intensification of the ring current will weaken the magnetospheric field earthward of the current system which is located between 3 and 7 Earth radii ($R_E$), pushing the cutoff latitudes equatorward. The same will apply to a strengthening of the cross-tail current, but on a different spatial scale as its inner edge is located approximately at 6–10 $R_E$.

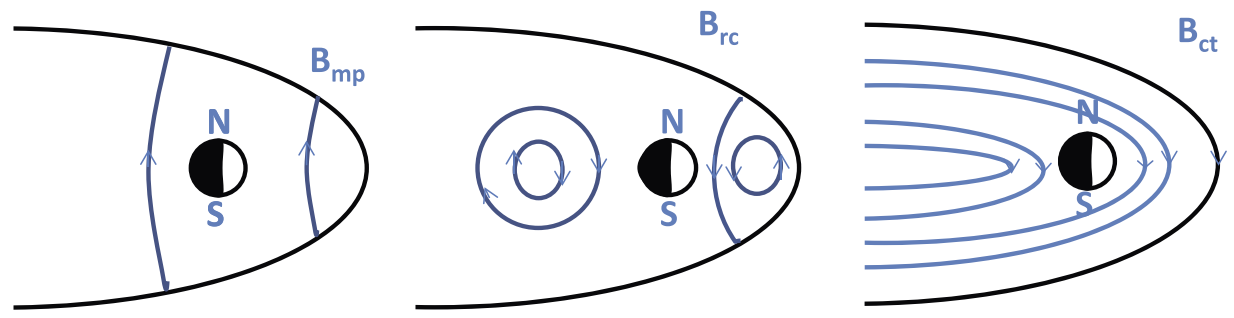

**Figure 1.** Simple sketch of the magnetic fields of the magnetopause current ($B_{mp}$), the ring current ($B_{rc}$), and the cross-tail current ($B_{ct}$).

The strength and location of the different current systems will vary with varying solar wind parameters. For example, an intensification of the solar wind pressure will increase the cross-tail current and the magnetopause currents which have opposite impact on the cutoff latitudes. The magnetopause current effect will be stronger on the dayside compared to the nightside and vice versa for the cross-tail current effect, resulting in a local time-dependent response of the cutoff latitude.

An adequate description of the time varying cutoff latitudes during SPEs has been the object for several studies applying trajectory-tracing techniques using the International Geomagnetic Reference Field and Tsyganenko models, and the global Lyon-Feder-Mobarry MHD simulation codes [*Blake et al.*, 2001; *Birch et al.*, 2005; *Smart et al.*, 2006; *Kress et al.*, 2010]. In addition to requiring extensive computing effort, the calculated cutoff latitudes are systematically poleward of the observed cutoff latitudes [*Dmitriev et al.*, 2010, and reference therein]. The discrepancies increase with increasing geomagnetic activity, which reflects the difficulties of predicting the magnetospheric field as it changes in both strength and configuration. Also, simpler empirical cutoff latitude models have been developed applying a linear dependence on either *Dst* or *Kp* [*Leske et al.*, 2001; *Birch et al.*, 2005; *Neal et al.*, 2013], a quadratic dependence of *Kp* [*Neal et al.*, 2013], and a function of multiple indices *Dst*, *Kp*, *AE*, and dipole tilt angle [*Dmitriev et al.*, 2010]. *Leske et al.* [2001] and *Neal et al.* [2013] both found a good correlation between observed cutoff changes and the *Kp* or the *Dst* changes taking into account a time-shift interval. *Neal et al.* [2013] found that the *Dst* changes occurred approximately 1 h after the cutoff changes, while the *Kp* changes preceded the cutoff changes by approximately 3 h making *Kp* favorable as a predictive model. However, *Neal et al.* [2013] disregarded periods when the magnetospheric field was impacted by a coronal mass ejection (CME) and the cutoff latitudes revealed strong day-night asymmetries. *Nesse Tyssøy et al.* [2013] showed that associated with a CME during the January 2012 SPE the dayside cutoff latitudes were pushed poleward while the nightside cutoff latitudes were pushed equatorward. Neither *Dst* nor *Kp* alone can reflect this dual mechanism.

Additionally, most of the theoretical and empirical models have mainly focused on high-energy proton (tens of MeV). However, protons less than 20 MeV are more affected by magnetic field changes and reveal a stronger day-night asymmetry, as well as stronger dawn-dusk asymmetries [*Fanselow and Stone*, 1972; *Dmitriev et al.*, 2010; *Nesse Tyssøy et al.*, 2013]. These protons are vital for understanding and quantifying the effect of energetic particle precipitation (EPP) on the middle atmospheric chemistry and dynamics. In studies concerning the EPP effect upon the atmosphere it is common to assume that the protons impact the polar atmosphere homogeneously above a fixed nominal boundary of for instance 60° geomagnetic latitude [e.g., *Jackman et al.*, 2009]. This is a gross oversimplification from the perspective of space physics. Consequently, the static model of the cutoff latitudes will miscalculate the distribution of particle energy deposition and the subsequent effects on chemistry and dynamics in the middle atmosphere [*Nesse Tyssøy et al.*, 2013].

To bridge the gap between the two science disciplines, we aim to provide a simple applicable cutoff latitude parameterization that includes both dayside and nightside cutoff variability. Our empirical basis is 6 selected SPEs from 2003 to 2012 with maximum particle flux unit (>10 MeV) above 1000 measured by Geostationary Operational Environmental Satellites (GOES) [http://umbra.nascom.nasa.gov/SEP/]. The magnetic field changes caused by the varying current sources are analyzed and discussed in context with observed particle latitude profiles from Polar Orbit Environment Satellites (POES). We assess the consequences of a strong day-night asymmetry upon the energy deposition distribution. Finally, the resulting cutoff latitude parameterization is utilized on the GOES particle fluxes and we evaluate its agreement with the energy deposition distribution estimated from the POES measurements.

## 2. Instrumentation and Data

Since 2002 POES 15, 16 and 17 were orbiting the Earth in polar, sunsynchronous orbits covering adequate local time intervals for analyzing both dayside and nightside cutoff variations. In 2005, 2006, and 2009 NOAA 18, METOP02 and NOAA 19 were launched into orbit further improving the time resolution and local time coverage. The different spacecraft are at around 850 km altitude with a period of approximately 100 minutes, and they are all carrying the Medium Energy Proton and Electron Detectors (MEPED) with the same nominal energy ranges. MEPED includes two proton solid-state telescopes that monitor the intensity of protons in six energy ranges from 30 to 6900 keV and ≥6900 keV pointing 9° and 89° to the local vertical. Additionally, MEPED includes an omnidirectional detector system, which covers a wide range of angles: 0° -60° from the vertical, for protons with energies 16–70 MeV. Combining the measurement from the vertical detector (pointing 9°) and the omnidirectional detector system we cover the proton energy range: 30 keV-70 MeV. The measurements are sorted into 1° latitude bins. We then obtain integral spectra applying monotonic piecewise cubic Hermite interpolating polynomials [*Fritsch and Carlson*, 1980] to the measured fluxes. The differential energy spectra are determined from the integral spectra. We define the cutoff location to be that invariant latitude where the count rate is half of its mean value above 70° CGM (Corrected GeoMagnetic) latitude in agreement with *Leske et al*. [2001].

**Table 1.** Six Moderate to Strong SPEs From 2003 to 2012 Giving the Particle Flux Unit, Maximum Pressure, and Minimum/Maximum *Dst*[a]

| Start (Day/UT) | Maximum (Day/UT) | Proton Flux (pfu >10 MeV) | Max Pressure (nPA) | Min *Dst* (nT) | Max *Dst* (nT) |
|---|---|---|---|---|---|
| [b]2 Nov 2003/2225 | 3 Nov/0600 | 1570 | 19.4 | −69 | 14 |
| 16 Jan 2005/0210 | 17 Jan/1750 | 5040 | 59.4 | −103 | 10 |
| 8 Sep 2005/0215 | 11 Sep/0425 | 1880 | 22.4 | −139 | 12 |
| 6 Dec 2006/1555 | 7 Dec/1930 | 1980 | 15.7 | −47 | −14 |
| 23 Jan 2012/0530 | 24 Jan/1530 | 6310 | 11.7 | −80 | 14 |
| [b]7 Mar 2012/0510 | 8 Mar/1115 | 6530 | 27.4 | −143 | 40 |

[a]Particle flux unit: http://umbra.nascom.nasa.gov/SEP/, *Dst*: WDC for Geomagnetism, Kyoto, Japan, Solar wind pressure: OMNIweb solar wind parameters.
[b]Do not cover entire event due to missing solar wind data or POES particle data.

## 3. Cutoff Latitude Variations With Geomagnetic Activity and Local Time

In general, it is only the strong SPEs (>1000 particle flux units) that are considered when estimating the effect of EPP upon the atmosphere. From 2003 to 2012 there were 9 events in this category [http://umbra.nascom.nasa.gov/SEP/]. We have selected six of these events, listed in Table 1, based on a requirement of good time coverage of the estimated cutoff latitudes on both the dayside and nightside, in addition to availability of solar wind measurements. The chosen events have different signatures in respect to the strength of the particle fluxes, solar wind pressure, and *Dst*.

Figure 2 shows the dayside (black) and nightside (blue) cutoff latitudes for 4 MeV and 16 MeV for all six events. The 4 MeV cutoff latitudes are located higher compared to the 16 MeV cutoff latitudes as the particles with higher energies (higher rigidities) penetrate deeper into the magnetosphere. The discrepancy is larger on the dayside, which implies a stronger day-night asymmetry in the cutoff latitudes for lower energies compared to higher energies. Both the 4 MeV and 16 MeV cutoff latitudes show large variability both on the dayside and nightside, but the variance is more pronounced on the dayside, resulting in a varying day-night asymmetry throughout the events.

The third panel for each event in Figure 2 shows the corresponding *Dst* and the pressure-corrected *Dst**, which is considered to be a more accurate measure of the magnetic field produced by the ring current [*Burton et al*., 1975]:

$$Dst^* = Dst - bp^{\frac{1}{2}} + c \qquad (1)$$

where $p$ is the ram pressure of the solar wind, $b = 15.8 \frac{\text{nT}}{\text{nPa}^{0.5}}$, and $c = 20$ nT is a correction factor related to the effect of magnetopause currents for average solar wind conditions. Although, there are numerous variants of the values of the constants $b$ and $c$ suggested in the literature, this first and frequently used set of constants is able to remove the positive deflection in the original *Dst* associated with large abrupt pressure increases unrelated to the ring current. The pressure itself is displayed in the fourth panel and the northward component of the interplanetary magnetic field (IMF), $B_z$, in the fifth panel. The sixth panel shows dawn to dusk solar wind electric field, $E_y$, which is the negative product of the solar wind speed and the northward component of the magnetic field, $B_z$. $E_y$ can be considered a proxy for the injection rate of the ring current when $-vB_z > 0.49$ mV, otherwise the injection rate is 0 [*Burton et al.*, 1975]. In other words, the ring current will only strengthen when $B_z$ points southward.

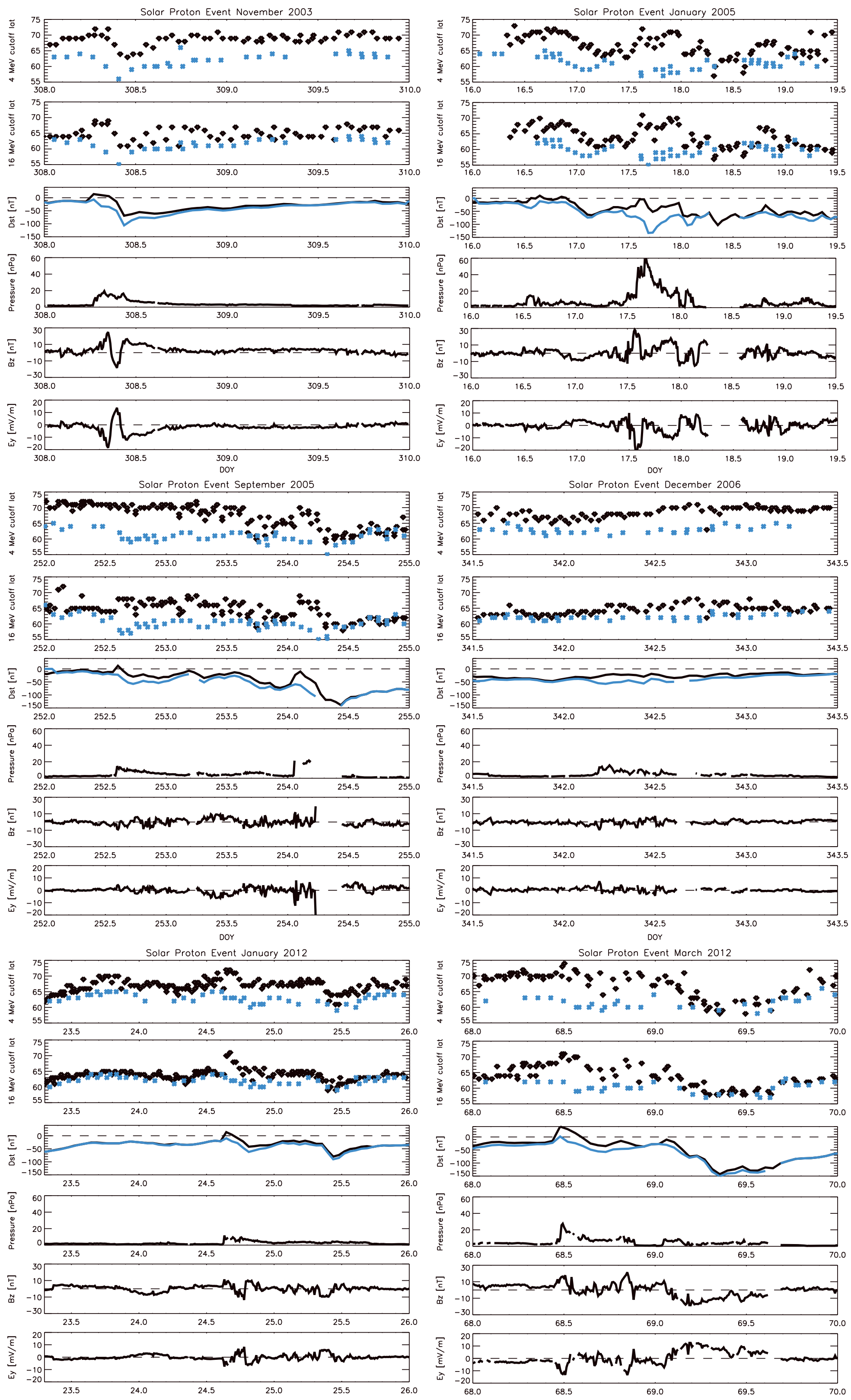

**Figure 2.** Six SPEs: (first and second panels) The 4 and 16 MeV cutoff latitudes estimated based on measurements from all available POES satellites for the dayside (black) and nightside (blue). (third panel) the corresponding *Dst* (black) and pressure-corrected *Dst** (blue). The time shifted (fourth panel) solar wind pressure, (fifth panel) interplanetary field strength of the $B_z$ coordinate (GSM), and (sixth panel) interplanetary electric field, $E_y$.

**Table 2.** Correlation Coefficients Between the Cutoff Latitudes and *Dst*, Pressure-corrected *Dst*, *Dst**, and Solar Wind Pressure for 4 and 16 MeV on Both Dayside and Nightside[a]

| Event | | Correlation 4 MeV Day | Correlation 16 MeV Day | Correlation 4 MeV Night | Correlation 16 MeV Night |
|---|---|---|---|---|---|
| SPE Nov 2003 | *Dst*: | 0.51 (S) | 0.50 (S) | 0.36 (S) | 0.30(S) |
| | *Dst**: | 0.61 (S) | 0.40 (S) | 0.63 (S) | 0.62(S) |
| | $\sqrt{p}$: | −0.29(S) | 0.07 (NS) | −0.74 (S) | −0.83(S) |
| SPE Jan 2005 | *Dst*: | 0.76 (S) | 0.85 (S) | 0.24 (S) | 0.16 (NS) |
| | *Dst**: | 0.41 (S) | 0.34 (S) | 0.72 (S) | 0.68 (S) |
| | $\sqrt{p}$ : | 0.28 (S) | 0.52 (S) | −0.62 (S) | −0.62 (S) |
| SPE Sep 2005 | *Dst*: | 0.88 (S) | 0.77 (S) | 0.31 (S) | 0.28 (S) |
| | *Dst**: | 0.82 (S) | 0.61 (S) | 0.43 (S) | 0.35 (S) |
| | $\sqrt{p}$ : | 0.22 (NS) | 0.46 (S) | −0.57 (S) | −0.67 (S) |
| SPE Dec 2006 | *Dst*: | 0.57 (S) | 0.55 (S) | 0.44 (S) | 0.53 (S) |
| | *Dst**: | 0.52 (S) | 0.22 (NS) | 0.67 (S) | 0.74 (S) |
| | $\sqrt{p}$: | −0.07 (NS) | 0.28 (S) | −0.23 (NS) | −0.27 (NS) |
| SPE Jan 2012 | *Dst*: | 0.76 (S) | 0.79 (S) | 0.43 (S) | 0.47 (S) |
| | *Dst**: | 0.71 (S) | 0.66 (S) | 0.65 (S) | 0.72 (S) |
| | $\sqrt{p}$: | 0.12 (NS) | 0.29 (S) | −0.50 (S) | −0.42 (S) |
| SPE Mar 2012 | *Dst*: | 0.86 (S) | 0.89 (S) | 0.08 (NS) | 0.42 (S) |
| | *Dst**: | 0.87 (S) | 0.86 (S) | 0.20 (NS) | 0.51 (S) |
| | $\sqrt{p}$: | 0.26 (NS) | 0.47 (S) | −0.49 (S) | −0.25 (NS) |

[a]S, significant: $p$ test $< 0.05$; NS, not significant: $p$ test $> 0.05$.

Focusing on September 2005 and January and March 2012 in Figure 2, it is clear that there are small differences between the 16 MeV dayside and nightside cutoff latitudes during periods of low-solar wind pressure levels (less than 2 nPa). The mean values of the latitude differences are 0.9°, 1.3°, and 1.8° for the three events, respectively. With elevated pressure levels (more than 2 nPa), the cutoff latitudes on the dayside are much higher than on the nightside, with mean values of the latitude differences 4.9°, 3.3°, and 4.7°, respectively. The change in the cutoff latitude often happens abruptly, in particular on the dayside as seen during, for example, the November 2003, September 2005, and January 2012 SPEs.

In general, both the dayside 4 MeV and 16 MeV cutoff latitudes correlate reasonably well with the *Dst* index within the 95% confidence interval as seen in Table 2. The correlation factors are particularly strong for the January and September 2005, and January and March 2012 SPEs with values ranging from 0.77 to 0.89 for 16 MeV cutoff latitudes. However, the same events show a low correlation between the *Dst* and the nightside cutoff latitudes with correlations of 0.16–0.47. The correlation is in particular poor in the time interval where *Dst* becomes less negative or positive as seen in Figure 2. A positive *Dst* suggests strong magnetopause currents strengthening the magnetic field on the dayside. The associated magnetic field changes are much weaker on the nightside. By applying the pressure-corrected *Dst*, removing the effect of the magnetopause currents, the correlation on the nightside improves for both 4 and 16 MeV cutoff latitudes as listed in Table 2.

In summary, the dayside and nightside cutoff latitudes show an opposite response to increased solar wind pressure which indicates different mechanisms dominating the dayside and nightside. The variety of interplanetary conditions is responsible for the complicated dynamics of the magnetospheric current systems, and subsequently the resulting cutoff latitudes. Ideally, to understand the asymmetric cutoff latitude variation, we need to understand the relative contributions of the different current systems and the solar wind conditions responsible for them. The following subsection focuses on the behavior of the dayside and nightside cutoff separately.

### 3.1. The Dayside Cutoff Variation

A cross-correlation analysis between the *Dst* and the cutoff latitudes shows that the 4 and 16 MeV dayside cutoff latitudes correlate best with the uncorrected *Dst* with no temporal offset to the cutoff latitude change as seen in Figure 3. This implies that the cutoff latitude variation is subject to the same current sources as the regular *Dst*. All of the SPEs, except December 2006, have short periods of positive *Dst* which suggests that strong magnetopause currents dominate over the negative contribution of the ring current. The magnetospheric field earthward of the magnetopause strengthens, and the cutoff latitudes are pushed poleward.

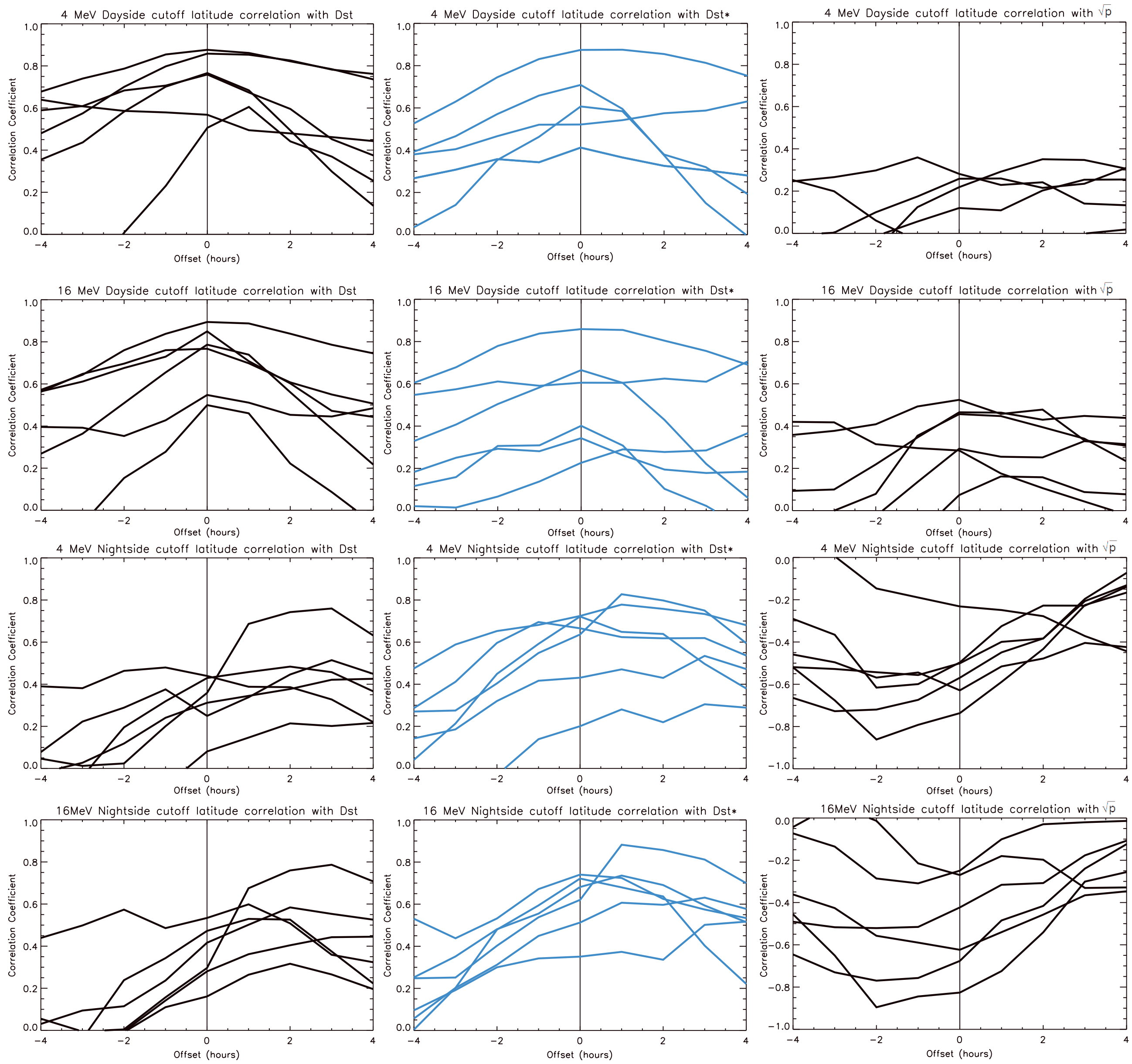

**Figure 3.** Time offset correlation dependence between 4 and 16 MeV cutoff latitudes and (left column) *Dst* (black lines), (middle column) *Dst** (blue lines), and (right column) square root of solar wind pressure (black lines) for both dayside and nightside. The *Dst*, *Dst**, and pressure are time shifted from 4 h before (−) to 4 h after (+) the cutoff latitude estimates.

The strengthening of the magnetopause currents is due to increased solar wind pressure. Still, there is not a strong correlation between the dayside cutoff latitudes and the pressure as seen in Table 2. This is confirmed by the cross-correlation analysis between solar wind pressure and dayside cutoff latitude shown in Figure 3. The result is somewhat surprising as the abrupt dayside increase in cutoff latitudes shown in Figure 2 clearly is associated with an increase in solar wind pressure. Looking at November 2003 and January 2012, the elevated pressure level is not enough to sustain the dayside elevated cutoff latitudes. The dayside cutoff latitudes start to move equatorward coincidental with the *z* component of the IMF turning negative. Another example is the January 2005 event. When the pressure increases, $B_z$ is positive, and the cutoff latitudes move poleward. In a short period where $B_z$ becomes negative the cutoff moves to lower latitudes, but when $B_z$ once again points northward, the cutoff latitude increases even though the pressure is decreasing.

**Table 3.** Regression Formulas That Best Reflect the Dayside Cutoff Latitude Variation in All Six SPEs for 1, 4, 8, 16, and 32 MeV, Including Their Average Correlation Factor With the Measured Cutoff Latitudes[a]

| Energy | Cutoff Latitude Parameterization | Correlation | Suggested Energy Range |
|---|---|---|---|
| 1 MeV | $0.090Dst + 0.02B_{ZN} + 72.0$ | 0.75 | $<$ 2 MeV |
| 4 MeV | $0.085Dst + 0.04B_{ZN} + 70.0$ | 0.72 | 2–6 MeV |
| 8 MeV | $0.065Dst + 0.14B_{ZN} + 68.5$ | 0.65 | 6–10 MeV |
| 16 MeV | $0.070Dst + 0.14B_{ZN} + 66.5$ | 0.73 | 10–20 MeV |
| 32 MeV | $0.055Dst + 0.10B_{ZN} + 64.5$ | 0.74 | $>$ 20 MeV |

[a]It also list the particle energy ranges applied on the energy estimates given in Figure 5.

There are potentially two physical links between the $B_z$ component and the cutoff variation. One is through the reconnection mechanism. A negative $B_z$ will allow the interplanetary and magnetospheric field to reconnect, resulting in a peel off of the dayside magnetic flux. In addition, a negative $B_z$ will support ring current growth through the particle injection, weakening the dayside magnetic field earthward of the ring current.

The strong correlation between the cutoff latitudes and the *Dst* suggests that some of the $B_z$ variation is already reflected in the *Dst* variation. However, the regression formulas that best reflect the dayside cutoff latitudes using only the *Dst* underestimate the latitude of the cutoff boundaries in periods of northward $B_z$. In periods of northward $B_z$ the dayside magnetic flux increases, making it harder for the solar wind protons to penetrate the closed field lines pushing the cutoff latitude poleward. Therefore, both the *Dst* and the northward $B_z$ value, $B_{z,N}$, are used in parameterization of the dayside cutoff latitudes. The optimizing routine is a least squares method. By stepwise changing the coefficients of these equations and summing the discrepancy between the modeled and measured cutoff latitudes for each event, the coefficients providing the smallest sum of the square of the residuals for all six events are chosen. The coefficient variation range is chosen such that the individual terms can impact the latitudes up to 15°. The regression formulas that best reflect the dayside cutoff latitudes, $\lambda_C$, for all six events are the following:

$$4\text{ MeV}: \ \lambda_C = 0.085Dst + 0.04B_{Z,N} + 70.5^\circ \quad (2)$$

$$16\text{ MeV}: \ \lambda_C = 0.070Dst + 0.14B_{Z,N} + 66.5^\circ \quad (3)$$

with the additional requirement of 58° CGM as a lower boundary for the dayside cutoff latitudes. The regression reveals an average correlation factor of 0.72 and 0.73, respectively. A more detailed energy resolution of the cutoff parameterization is given in Table 3. Figure 4 shows the resulting cutoff parameterization. The parameterization reflects the main dayside cutoff levels very well. The good correlation between the dayside cutoff latitudes and the *Dst*, even when the *Dst* turns positive, suggests that the main sources for changing the dayside cutoff latitude boundaries are the ring current and the magnetopause currents. In addition, the varying dayside magnetic flux due to magnetic reconnection will also play a role.

### 3.2. The Nightside Cutoff Variation

The cross-correlation analysis between the regular and pressure-corrected *Dst* and the nightside cutoff latitudes in Figure 3 shows that the correlation maximizes for both the *Dst* and *Dst** + 1–3 h offset to the 16 MeV cutoff latitude change. That means that the cutoff variations lead the corresponding change in *Dst*, which implies that the main cause for the nightside cutoff latitudes probably is not the ring current itself as the cutoff changes occur prior to the ring current buildup/loss. This feature is also noticeable in Figure 2 for November 2003 where the decrease in the nightside cutoff (day of year (DOY) 308.2) occurs prior to the corresponding *Dst* decrease. The interpretation is supported by the interplanetary electric field which indicates little or low ring current injection rate. The same is evident for the January 2012 SPE (DOY 24.6). When the main solar wind pressure increase occurs, the interplanetary electric field does not support ring current injection. Yet the nightside cutoff latitude is pushed equatorward and the magnitude of the pressure-corrected *Dst* increases, which supports the buildup of the tail current due to the solar wind pressure increase. The buildup and recovery of the tail current have a shorter response time to the changes in the interplanetary medium compared to the buildup and recovery of the ring current. Also, as the tail current is located further from the Earth compared to the ring current, it will have a weaker signature on the ground, but it will still have a significant impact on cutoff latitudes in particular for the lower energies.

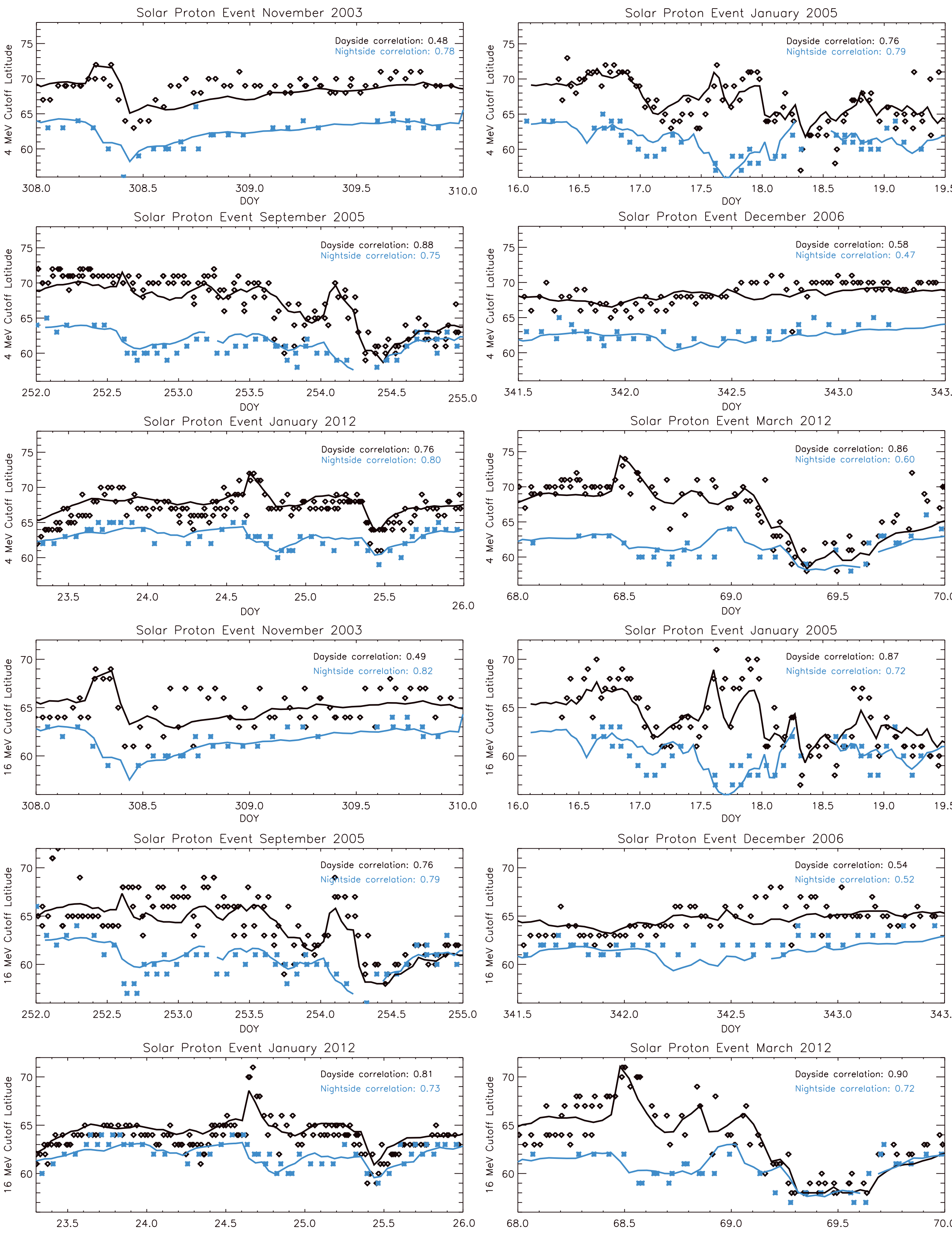

**Figure 4.** Six SPEs: The 4 and 16 MeV cutoff latitudes estimated based on measurements from all available POES satellites for the dayside (black diamonds) and nightside (blue asterisks). The black and blue solid lines shows the dayside and nightside cutoff parameterization determined from equations (2)–(5).

In other words, the time shift suggests that neither the regular nor the pressure-corrected *Dst* sufficiently includes the physical mechanism responsible for the nightside cutoff latitude changes. Table 2 shows that the cutoff latitudes have a correlation of up to −0.83 with the square root of the solar wind pressure in November 2003, while in January 2012 it was only −0.42. Cross-correlation analysis of the square root of the solar wind pressure with the nightside cutoff latitude shows that the solar wind pressure leads the cutoff latitude by up to 2 h prior to the cutoff change. The negative correlations imply that the impact is mainly through the tail current and not the magnetopause current. However, the apparent offset might be understood considering that the pressure is responsible for enhancing both the magnetopause current and the tail current, which have opposite impact on the cutoff latitude. It is evident from the dayside cutoff latitude variation that the positive deflection in the *Dst*, caused by strengthening the magnetopause currents, has no offset in time. The positive impact will be weaker on the nightside, but still present in the pressure cross-correlation analysis. For example, looking at the SPE September 2005 (DOY 252.7), the nightside cutoff latitude is pushed equatorward quite abruptly in line with the pressure with no apparent offset in time.

The offset and the strength of the correlation appear to be more prominent for 4 MeV compared to 16 MeV, which could be attributed to the tail current being localized at a distance from the Earth that will give a stronger impact when the cutoff are located at higher latitudes. For the 16 MeV cutoff latitudes both the offset and strength of the correlation are in general weaker and more variable compared to 4 MeV.

The varying correlation from one SPE to another, in particular for 16 MeV cutoff latitudes, could be symptomatic of how important the pressure-related tail current buildup is compared to the ring current in the different storms. A nightside cutoff parameterization should therefore reflect both the ring current strength and the tail current in the region of the magnetosphere responsible for the associated field changes. Based on these criteria we introduce a new pressure-dependent term representing the effect of the tail current on the cutoff latitudes: $A * \text{pressure}^B$. By stepwise changing the *A* and *B* values in a least squares method, we find which factors that best describe the nightside cutoff latitudes for all six SPEs with a potential pressure offset in time up to 2 h.

The third root of the pressure gave the best result for both 4 and 16 MeV with no offset in time but with stronger pressure dependence for 4 MeV. The corresponding regression for the cutoff latitudes is

$$4\text{ MeV}: \ \lambda_C = 0.040Dst - 3.2p^{\frac{1}{3}} + 68.5 \tag{4}$$

$$16\text{ MeV}: \ \lambda_C = 0.035Dst - 3.0p^{\frac{1}{3}} + 67.0 \tag{5}$$

with the additional requirement of 56° CGM latitude as the lower boundary. Since the cutoff latitudes are much more confined on the nightside compared to the dayside, finer energy steps are not considered necessary. The regression fits well with the general cutoff latitude variations as seen in Figure 4. The average correlations for 4 MeV and 16 MeV are 0.70 and 0.72, respectively.

The new cutoff parameterization differs from the pressure-corrected *Dst* in the sense that it not only removes the positive contribution from the magnetopause currents but it also further emphasizes the link between the pressure and tail current buildup. The respective correlations are not time shifted as seen in the pressure-corrected *Dst*, since the tail current has a fairly short response time to the pressure. The suggested empirical parameter is not meant to reflect the relative contribution of the ring current and tail current to the *Dst*. It should only be used in the estimation of nightside cutoff latitudes.

## 4. Asymmetric Energy Deposition

The cutoff latitude parameterization represents the day-night asymmetry in cutoff latitudes in a simplistic manner so it could readily improve the models of energetic particle precipitation used in atmospheric studies during SPEs. Frequently, GOES particle measurements are used assuming uniform proton precipitation above a fixed nominal latitude boundary. Comparing the energy deposition derived from POES and the GOES satellites, *Nesse Tyssøy et al.* [2013] showed that this assumption would give an overestimate of the energy deposition on the dayside below 67° CGM latitudes by 50–100% at 70 km in the main phase of the January 2012 event. Based on the cutoff latitude variation shown in Figure 2, the January 2012 event is by no means an extreme event considering the day-night asymmetry. In fact, most of the events have stronger and longer-lasting day-night asymmetries which emphasize the need of taking the cutoff variation throughout an event into consideration when evaluating the energy deposition.

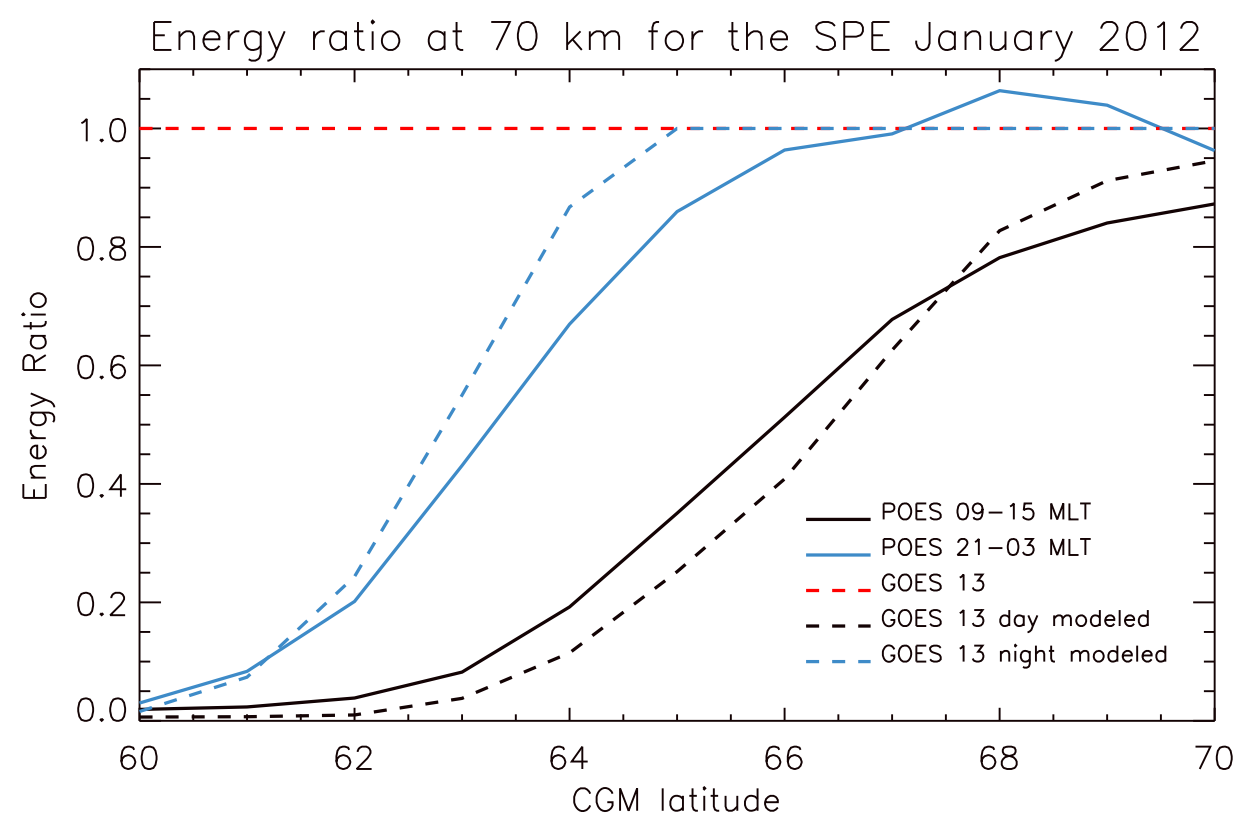

**Figure 5.** The estimated energy deposition at 70 km from the POES measurements for day (black) and night (blue) relative to the energy deposition estimated from GOES 13 measurements for the entire SPE January 2012. The red dashed line is the energy deposition from GOES 13 normalized to 1.0, while the black and blue dashed lines are the energy deposition when applying the dayside and nightside cutoff parameterization to the GOES fluxes.

Figure 5 shows the estimated ratio of the energy deposition at 70 km altitude based on the POES measurements relative to the energy deposition estimated from GOES 13 for day (black) and night (blue) for the entire SPE January 2012. The energy deposition height profile for protons is calculated based on range energy of protons in air given by *Bethe and Ashkin* [1953]. The atmospheric densities are retrieved from the MSIS-E-90 model [*Hedin*, 1991]. We have assumed that the proton fluxes are isotropic over the downward hemisphere. The GOES measurement overestimates the energy deposition by 25% already at 68° CGM latitude on the dayside and at 64° CGM latitude on the nightside. However, by applying the above given parameterization on GOES particle fluxes, we could estimate new particle fluxes representative for different latitudes. For a specific energy interval we can apply the appropriate cutoff parameterization and modify the respective GOES fluxes accordingly:

1. parameterized cutoff latitude < latitude: modeled flux = GOES flux;
2. parameterized cutoff latitude > latitude: modeled flux = 0; and
3. parameterized cutoff latitude ~ latitude: modeled flux = GOES flux/2.

The last criterion is in accordance with the definition used to estimate the cutoff latitudes in this study. The model should not be used in the few cases when the cutoff latitude boundary is above the foot point of the GOES satellite. Looking at fluxes from individual passes of POES satellites, we find that the flux starts to decrease quite steeply when we reach the cutoff latitude, making it reasonable to assume a zero flux at lower latitudes. The effect on the energy deposition is shown as dashed lines for the modified dayside (black) and nightside (blue) energy deposition, respectively. (For the nightside we have used equation (4) for proton fluxes < 12 MeV and equation (5) for proton fluxes > 12 MeV. The applied energy ranges for the dayside are given in Table 3.) The modified GOES energy deposition now capture the day-night asymmetry and is in good agreement with the energy deposition predicted from the POES measurements for the entire event.

To avoid discontinuity between day and night in a study or applied to a model, one can, for example, use an interpolation routine for the cutoff latitude boundaries for morning and evening to get a smooth transition. This simple day-night division might, however, mask some features related to dawn-dusk asymmetries as reported by *Dmitriev et al.* [2010] and reference therein. Focusing on the January and March SPEs in 2012, where six POES/MEPED satellites were in operation enabling us to study the magnetic local time (MLT) variation of the cutoff latitudes in more detail, we find some evidence (not shown here) that dawn-dusk asymmetry can be taken into account considering the direction of the solar wind. As we consider the solar wind pressure and the subsequent magnetopause current as the most likely cause for the poleward cutoff latitude push on the dayside, the local time where we find the maximum strength of this current on the magnetopause will vary with the pressure angle. This mechanism may be further investigated with the more recent and future SPEs where the POES satellite coverage is better. However, we consider this to be out of the scope of this paper as our purpose is to provide an easily applicable cutoff latitude parameterization that takes into account the main features of the cutoff latitude variation.

## 5. Conclusion

Based on six moderate to strong SPEs from 2003 to 2012, we investigate the variation of the proton cutoff latitudes using measurement from all available POES satellites. We find a strong storm time variation of the cutoff latitudes and an associated day-night asymmetry which have consequences for the distribution of the proton energy deposition and subsequently the effects on the chemistry and dynamics in the middle atmosphere. Considering the physical mechanisms influencing the cutoff latitudes, we provide a parameterization for both the dayside and nightside using only the *Dst*, the northward component of the interplanetary magnetic field, and solar wind pressure. The parameterization could also be utilized on the GOES particle fluxes, and the resulting energy deposition will improve the quantification of the total energy being deposited and capture the related day-night asymmetry during SPEs.

**Acknowledgments**
This work was supported by the Research Council of Norway grants 223252/F50 and 184701. We acknowledge NOAA, Boulder, USA, for GOES and POES energetic particle data (http://www.ngdc.noaa.gov/), WDC for Geomagnetism, Kyoto, Japan, for *Dst* indices (http://wdc.kugi.kyoto-u.ac.jp/), and SPDF Goddard Space Flight Center for solar wind parameters (http://omniweb.gsfc.nasa.gov/form/omni_min.html).

Michael Balikhin thanks the reviewers for their assistance in evaluating this paper.